\documentclass{article}
\usepackage{url}
\usepackage{graphicx}
\usepackage{calligra}
\usepackage[T1]{fontenc}
\usepackage{float}
\restylefloat{table}
\begin{document}
\vspace*{-3cm}
\begin{flushright}
\today

\end{flushright}


\vspace{5.0ex}

\begin{center}
  
\begin{Large}
  {\bf DISCOVERING THE SIGIFICANCE OF 5$\sigma$ }
  
\end{Large}

\end{center}

\begin{center}

\vspace{4.0ex}

\vspace{1.0ex} {\Large Louis Lyons}\\
\vspace{1.0ex}
\emph{Blackett Lab., Imperial College, London SW7 2BW, UK} \\ 
and\\
\emph{Particle Physics, Oxford OX1 3RH, UK} \\
\vspace{2.0ex}
e-mail: l.lyons@physics.ox.ac.uk
\vspace{4.0ex}

ABSTRACT

\end{center}
We discuss the traditional criterion for discovery of requiring a significance corresponding to 5$\sigma$; and whether
a more nuanced criterion might be better.
\vspace{0.3cm}


\section{Introduction}
It has become the convention in Particle Physics that in order to claim a discovery of some form of 
New Physics, the chance of just the background having a statistical fluctuation at least as large as the 
observed effect is equivalent to the area beyond 5$\sigma$  in one tail of a normalised Gaussian distribution, or smaller; 
i.e. the $p$-value is no larger than $3\times 10^{-7}$. Indeed, journals are very reluctant to allow the word
`discovery' to appear in the title, abstract or conclusions of a paper, unless the $p$-value is that small.
In this short note, we list the arguments for and against this attitude.

\section{Why 5$\sigma$?}
Statisticians who hear about the 5$\sigma$ criterion are very skeptical about it being sensible. In 
particular, they claim that it is very uncommon that statistical probability distributions accurately describe
reality so far out in the tails. Particle Physics does differ from other fields in that we do believe that in many 
situations the numbers of observed events do follow a Poisson distribution, and so small tail probabilities
may well be meaningful. However, certainly as far as systematic effects are concerned, even in Particle
Physics there is often a degree of subjectivity in specifying the exact magnitude of the effect, let alone believing
in its assumed distribution far into the tails of the observed effect. Thus the Statisticians' criticism may well be valid.

The traditional arguments for the 5$\sigma$ criterion are:
\begin{itemize}
\item{History: 

In the past there have been many `phenomena' that corresponded to 3 or 4$\sigma$ effects that have 
gone away when more data were collected. The 5$\sigma$ criterion is designed to reduce the number 
of such false claims. However, it may well be worth checking whether the fraction of false claims at the 5$\sigma$
level is indeed smaller than at 4$\sigma$.   }

\item{Look Elsewhere Effect (LEE):

The significance of an interesting peak in a mass spectrum is defined in terms of the probability of a background 
fluctuation producing an effect at least as large  as the one actually seen. If the background fluctuation is 
required to be at the observed mass, this probability is called the local $p$-value. If, however, the restriction 
on the location of the background fluctuation is removed, this is instead the global $p$-value, and is larger than the local 
one because of the LEE. 

Ambiguity in the global $p$-value arises because of the choice of requirements that are imposed 
on the location\footnote{If the width and/or shape of the possible signal are regarded as free parameters,
they also contribute to the LEE factor.}
 of the background 
fluctuation. Possibilities include (a) any reasonable mass; (b) anywhere in the current analysis; (c) anywhere in 
analyses by members of the experimental Collaboration; etc. In addition to the local $p$-value, it  is desirable 
to quote at least the global one corresponding to option (a). It is clearly important to decide in advance of 
the analysis what procedure will be used; and to specify in any publication exactly how the 
global $p$-values are defined.
}

\item{Subconscious Bayes' Factor:    

When searching for a discovery, the data statistic that is used to discriminate between just background
(known as the null hypthesis $H_0$) and 
`background plus signal' ($H_1$) is often the likelihood ratio $L_1/L_0$ for the two hypotheses; and 
the 5$\sigma$ criterion is applied to the observed 
value of this ratio, as compared with its expected distribution assuming just background. However, more relevant to the discovery claim is the ratio of probabilities for the two hypotheses. These are related by
Bayes Theorem:
\begin{equation}
\frac{P(H_1|data)}{P(H_0|data)}  =   \frac{p(data|H_1)}{p(data|H_0)}  \times  \frac{\pi_1}{\pi_0}
\end{equation}
where $p(data|H_i)$ and $\pi_i$ are respectively the likelihood and the prior probability for hypothesis $H_i$;
and $P(H_1|data)$ is the hypothesis' posterior probability. So the ratio of probabilities we assign to 
the hypotheses is their likelihood ratio times their prior probability ratio. Hence, in this approach, even if the likelihood 
ratio favours $H_1$, we would still prefer $H_0$ if our prior belief in $H_1$ was very 
low. An example would be that, in order to claim that we had discovered energy non-conservation
in proton-proton collisions at the LHC, we would require extremely strong evidence from the data because
our prior belief in energy non-conservation is very low. Similarly, if an experiment seemed to show that neutrinos travel 
faster than light, we might well have a tendency to believe that there was some  
undiscovered systematic, rather than that neutrinos were a potential source for violating causality. 

The above argument is clearly a Bayesian application of Bayes Theorem, while analyses in Particle Physics 
usually have a more frequentist flavour.  Nevertheless, this type of reasoning does and should play a role in
requiring a high standard of evidence before we reject well-established theories. There is sense to the 
oft-quoted maxim `Extraordinary claims  require extraordinary evidence'. } 

\item{Systematics:  

It is in general more difficult to estimate systematic uncertainties than statistical ones. Thus a n$\sigma$
effect in an analysis where the statistical errors are dominant may be more convincing that one where the 
n$\sigma$ claim is dominated by systematics errors. Thus in the latter case, a 5$\sigma$ claim should be 
reduced to  merely 2.5$\sigma$ if the systematic uncertainties 
had been underestimated by a factor of 2;  this corresponds 
to the $p$-value increasing from $3\times10^{-7}$ by a dramatic factor of $2 \times10^4$. 
The current 5$\sigma$ criterion   is partially motivated by a cautious approach
to somewhat vague estimates of systematic uncertainties. 

Bob Cousins has recommended that for systematic-dominated analyses, it is desirable to assess by how much the systematics must be increased in order to reduce a significance above 5$\sigma$ to that level; and 
then deciding whether this level of systematic effect is plausible or not. }
\end{itemize} 

\begin{table}
\begin{center}
\caption{Summary of some searches for new phenomena, with suggested numerical values for the number of
$\sigma$ that might be appropriate for claiming a discovery. }
\begin{tabular}{|c|c|c|c|c|c|}
\hline
Search   &    Degree of    &  Impact   & LEE    &    Systematics  &   Number          \\
              &        surprise   &                &           &                         &   of $\sigma$     \\

\hline \hline
Higgs search  &     Medium   &    Very high   &      Mass &      Medium    &      5   \\  \hline      
Single top     &     No     &       Low    &      No      &    No     &     3   \\  \hline  
SUSY    &     Yes     &      Very high     &       Very large     &     Yes     &   7     \\    \hline
$B_s$ oscillations    &   Medium/low    &        Medium      &     $\Delta m $    &     No   &   4   \\    \hline 
Neutrino oscillations   &     Medium     &       High     &   $sin^2(2\theta), \Delta m^2$   &    No &   4   \\ \hline 
$B_s \rightarrow \mu\mu$   &    No      &     Low/Medium   &    No    &       Medium     &    3      \\   \hline
Pentaquark   &      Yes      &    High/very high    &       M, decay mode     &     Medium   &   7    \\ \hline
$(g-2)_{\mu}$ anomaly   &   Yes     &    High     &    No    &   Yes     &   4    \\    \hline
H spin $\neq$  0    &      Yes    &      High     &    No    &  Medium    &     5    \\   \hline    
$4^{th}$ generation $q, l, \nu$  &     Yes  &    High   &     M, mode   &    No   &    6   \\   \hline
v$_\nu$ > c   &   Enormous   &    Enormous   &      No     &   Yes    &     >8  \\ \hline
Dark matter (direct) &   Medium    &    High    & Medium      &     Yes    &    5  \\ \hline
Dark energy   &     Yes     &    Very high     &      Strength      &      Yes      &     5    \\ \hline
Grav waves    &     No     &     High   &    Enormous      &     Yes       &    7    \\    \hline      
\end{tabular}
\end{center}
\end{table}

\section{Why not 5$\sigma$?}

There are several reasons why it is not sensible to use a uniform criterion of 5$\sigma$ for all searches for new 
physics. These include most of the features that we included as supporting the use of the 5$\sigma$ criterion. 

\subsection{LEE}
This  can vary enormously from search to search. Some experiments are specifically designed to measure one parameter, which is sensitive to new physics, while others use general purpose detectors which can produce a whole range of potentially interesting results, and hence have a larger danger of  a statistical fluctuation somewhere in the background. 
An example of an experiment with an enormous LEE is 
the search for gravitational waves; these can have a variety of different signatures, occur at any time 
and with a wide range of frequencies, durations, etc.
 
It is illogical to penalise a single-measurement analysis simply because others have a significant LEE.

\subsection{Subconscious Bayes' Factor}
Some experiments search for effects which may be 
very weak but which are predicted to occur in the Standard Model (S.M.), while others may be 
very speculative (e.g. neutrinos 
travelling faster than the speed of light; search for mini-Black Holes at the LHC;.....). Clearly, it is unreasonable to to 
require the same level of significance for both types of search. 

\subsection{Systematics}
Some experiments are very sensitive to systematics, while in others it plays a 
minor role. Thus attempts to discover substructure of quarks by  looking at dijet events produced in quark-quark 
scattering at high enegy hadron colliders are dependendent on the parton distribution functions (pdf's) for the 
initial state beam hadrons. An over-optimistic estimate of the accuracy of pdf's in the relevant kinematic region can then 
result in a significant-looking signature of quark substructure, but which would be much less significant when 
more realistic pdf uncertainties are allowed for. 

It is far better to have a procedure which allows in some way for uncertain systematics than to raise the significance level 
for all experiments as a way of coping with this.

\subsection{Summary of some experimental searches}
Table 1 provides a summary of some experimental searches for new physics. The topics listed all correspond to 
searches that have actually been performed; some of these have resulted in actual discoveries, others have 
made claims which have subsequently been withdrawn, while others have set limits on the sought-for effect.
The remarks in the Table, however, are not related to any specific experiment, but are supposed to be more generic.
The final column contains a set of suggestions for the significance level required, after making some attempt 
to correct for the LEE, before a discovery could be claimed.  The spirit of these numbers is to provoke 
discussion of this issue (rather than being a rigid set of rules), by suggesting a graded set of significance 
levels for different types of experiments.  

We briefly contrast the various effects that are important for some of the searches in Table 1:
\begin{itemize}

\item{ $(g-2)_{\mu}$:

This experiment, designed to measure the anomolous magnetic moment of the muon, basically determines just this
quantity, and so there is essentially no LEE. Almost by definition, if there is no physics beyond the S.M., the 
experimental value should agree with the S.M. prediction, so any significant deviation would be an indication of new physics.
}
\item{Single top:

The top quark was discovered in $t \bar{t}$ pair production in proton-antiproton collisions at the Tevatron. The S.M. 
also predicts a somewhat smaller production rate for single top-quark production. This was searched for, and also found.
Since the mass of the top quark was known from the pair-production discovery mode, there was no LEE due to unknown mass here. Also, the Subconscious Bayes Factor here actually {\bf favours} single top production, since it is predicted 
by the S.M., and indeed it would be very surprising if it were not produced or was highly surpressed. Thus this 
seems to be a good case where 5$\sigma$ should not have been  a requirement before confimation of this production  
mode was accepted. 
}
\item{Higgs boson:

Even before the discovery of a Higgs boson at CERN's LHC, it was widely expected to exist (although
it would not have been completely a surprise if it did not). Thus a discovery claim was not a surprise, but 
because of its very important role in giving mass to the other elementary particles, its impact was large. The LEE 
was due to the unpredicted Higgs mass, but there were no further factors from the various possible decay modes,
whose branching ratios were predicted by the S.M. 
}

\item{Supersymmetry (SUSY):

It would have been very satisfying if SUSY had reduced the number of fundamental particles by pairing each known fermion
with a known boson. This did not happen, and so SUSY has essentially doubled the number of basic particles, by 
hypothesising SUSY partners for each known particle. Thus there are many different types of SUSY particles to search 
for and so, combined with a variety of possible mases and decay modes, there is a large SUSY LEE factor. Combined 
with the high impact of discovering a whole new sector of fundamental particles, the level of significance for a SUSY
discovery claim should be high. 
} 
\item{Mini-black holes:

There have been suggestions  that mini-Black Holes could perhaps be produced in the high energy proton-proton
 collisions at the LHC.
These are highly speculative particles, with a large LEE factor arising from their unknown mass and decay modes. A high standard of significance should be required for a claim of their discovery.
}
\end{itemize}

\subsection{Jeffreys-Lindley Paradox}
A criticism of a fixed level of $p$-value for discovery comes from the Jeffreys-Lindley
(J-L) paradox\cite{J_L_paradox}. 
Basically, this draws 
attention to the way a fixed significance level for discovery does not cope with a situation where the 
amount of data is increasing. An example involves testing a simple hypothesis $H_0$ (e.g. a Gaussian
 centred at $\mu = 0$) against a composite one (e.g. $\mu > 0$). It turns out that for a wide variety of 
priors for $\mu$, the Bayesian posterior probabilities for the hypotheses can favour $H_0$, while the frequentist 
$p_0$-value can reject $H_0$.

A simplified version of this effect is illustrated in Table 2; this uses simple hypotheses for both $H_0$ and $H_1$,
and does not require priors. 
This example involves  a counting 
experiment where in the first run 10 events are observed, when the null hypothesis $H_0$ predicts 1.0 
and the alternative $H_1$ predicts 10.0; both the $p$-value for the null hypotheses ($p_0$) and the likelihood 
ratio disfavour $H_0$. 
Then the running time is increased by a factor of 10, so that the expected numbers according to $H_0$ 
and $H_1$ both increase by a factor of 10, to 10.0 and 100.0 respectively. With 31 observed events, 
$p_0$ corresponds to about 
5$\sigma$ as in the first run; but despite this the likelihood ratio now strongly favours $H_0$. This is simply because the 
5$\sigma \ n_{obs}$ = 10 in the first run was exactly the expected value for $H_1$, but with much more data
the  5$\sigma \ n_{obs}$ = 31 is way below the $H_1$ expectation. This example thus shows that for a 
fixed $p$-value such as $3 \times 10^{-7}$  the likelihood-ratio can favour either hypothesis.

The conclusion from the J-L paradox is that the $p_0$  cut used to reject $H_0$ should decrease with increasing amount
of data. However, there appears to be no obvious way of implementing this, and Particle Physics tends to use
fixed levels of cuts, independent of the data size. 

A discussion of the J-L Paradox in its application to Particle Physics will be found in ref. \cite{Cousins_on_Lindley}.

\section{Conclusions}

It would be very useful if we could distance ourselves from the attitude of  `Require 5$\sigma$ for all discovery claims'.
This is far too blunt a tool for dealing with issues such as the Look Elsewhere Effect, the plausibility of the 
searched-for effect, the role of systematics, etc., which vary so much from experiment to experiment. 
The problem is to produce an agreed alternative. Table 1 is an attempt to stimulate discussion about
a graded set of significance levels for different types of experiments.   

A possibility would be to replace the significance hurdle, which results in a binary output\footnote{`Yes we achieved the necessary significance level to claim a discovery' or `Oh dear, we missed it'.}, by simply quoting the observed 
local and where relevant global $p$-values. This would make the distinction between 
4.9$\sigma$ and 5.0$\sigma$ much less crucial, and it would 
also show up the difference between 5.0$\sigma$ and 8.0$\sigma$. 
Unfortunately, however, this may  not be acceptable, as the general feeling 
 may well be to maintain emphasis on a fixed criterion to answer the question 
`Is this a discovery or not?'

\vspace{0.2in}

I wish to thank members of the Statistics Committes of the CMS and CDF Collaborations (especially Bob 
Cousins and Tom Junk respectively) for enlightening and useful discussions on this and many other
statistical issues; and my colleagues on these experiments for asking many thought-provoking questions.

\begin{table}
\begin{center} 
\caption{Comparing $p$-values and likelihood ratios}
\begin{tabular}{|c|c|c|}
\hline
   &   First data set    &     Second data set    \\ 
\hline    \hline
$H_0$     &    Poisson, $\mu=1.0$      &     Poisson, $\mu=10.0$    \\ \hline
$H_1$     &    Poisson, $\mu=10.0$    &     Poisson, $\mu=100.0$     \\    \hline
$n_{obs}$ &      10                        &    31 \\     \hline
$p_0$     &  $1.1 \times 10^{-7}$   &    $0.8 \times 10^{-7}$    \\
              &     5.2$\sigma$    &      5.3$\sigma$      \\     \hline
$L_0/L_1$ &     $8 \times 10^{-7}$ &     $1.2  \times 10^{+8}$     \\
               &  Strongly favours $H_1$ &   Strongly favours $H_0$  \\   \hline
\end{tabular}
\end{center}
\end{table}



%

%
\end{document}